\documentclass[fleqn,10pt]{wlscirep}
\usepackage[utf8]{inputenc}
\usepackage[T1]{fontenc}
\usepackage{bm}

\newcommand{\Prob}{\operatorname{P}}

\newcommand{\given}{\operatorname{|}}
\newcommand{\X}{\mathbf{X}}
\newcommand{\D}{\mathcal{D}}
\newcommand{\G}{\mathcal{G}}
\newcommand{\M}{\mathcal{M}}
\newcommand{\LL}{\mathcal{L}}

\newcommand{\PXi}{\Pi_{X_i}}
\newcommand{\XPi}{X_i \given \PXi}
\newcommand{\T}{\Theta_{X_i}}

\newcommand{\BIC}{\mathrm{BIC}}
\newcommand{\DAG}{\mathrm{DAG}}

\title{The Probabilistic Backbone of Data-Driven Complex Networks: An example in Climate\footnote{This is a pre-print of an article published in Scientific Reports. The final authenticated version is available online at: https://doi.org/10.1038/s41598-020-67970-y.}}

\author[1,*]{Catharina E. Graafland}
\author[1]{Jos\'e M. Guti\'errez}
\author[1]{Juan M. L\'opez}
\author[1]{Diego Paz\'o}
\author[1]{Miguel A. Rodr{\'\i}guez}
\affil[1]{Instituto de F{\'\i}sica de Cantabria, 
	CSIC--Universidad de Cantabria, Avenida de Los Castros, 
	E-39005 Santander, Spain}

\affil[*]{catharina.graafland@unican.es}

\begin{abstract}
Complex systems often exhibit long-range correlations so that typical 
observables show statistical dependence across long distances. These 
teleconnections have a tremendous impact on the dynamics as 
they provide channels for information transport across the system 
and are particularly relevant in forecasting, control, and data-driven
modeling of complex systems. These statistical interrelations among 
the very many degrees of freedom are usually represented by the 
so-called correlation network, constructed by
establishing links between variables (nodes) with pairwise correlations 
above a given threshold. Here, with the climate system as an example, we 
revisit correlation networks from a probabilistic perspective and show 
that they unavoidably include much redundant information, resulting in overfitted 
probabilistic (Gaussian) models. As an alternative, we propose here the use of more 
sophisticated probabilistic Bayesian networks, developed by the machine 
learning community, as a data-driven modeling and prediction tool. 
Bayesian networks are built from data including only 
the (pairwise and conditional) dependencies 
among the variables needed to explain the data ({\it i.e.}, maximizing the 
likelihood of the underlying probabilistic Gaussian model). This results in 
much simpler, sparser, non-redundant, networks still encoding the complex structure of the dataset as revealed by standard complex measures. Moreover, the networks are capable 
to generalize to new data and constitute a truly 
{\em probabilistic backbone} of the system. When applied to climate data, it is shown that Bayesian networks faithfully reveal the various long-range teleconnections relevant in the dataset, in particular those emerging in El Niño periods.
\end{abstract}
\begin{document}

\flushbottom
\maketitle
%
%
\thispagestyle{empty}

\section*{Introduction}

Due to the widespread interest of the scientific community in 
data science, an increasing body of 
research in the field of complex networks is now focusing on the development 
of graph machine learning algorithms for analysis and prediction~\cite{battaglia_relational_2018}. 
Some relevant applications include link prediction~\cite{mutlu_review_2019,varghese_machine_nodate}, 
network embedding~\cite{cui_survey_2017}, 
pattern mining ~\cite{karunaratne_learning_2014},  
and graph neural networks~\cite{Scarselli_graph_2009}. 
Most of this research is currently oriented towards the application of deep learning methods~\cite{cao_deep_2016,niepert_learning_2016,kipf_semi-supervised_2017,zitnik_deep_2019}. 
However, there are a number of traditional machine learning methods that
could be used in the modern context of complex graph analytics and prediction. 
For instance, Bayesian network (BN) models~\cite{castillo_expert_1997} are a 
sound and popular machine learning technique used to build tractable 
probabilistic models from data using auxiliary graphs--- representing the 
most relevant (pairwise and conditional) dependencies among the variables 
needed to explain the data as a whole (maximizing the likelihood of the underlying 
Gaussian model). These models have been successfully applied in a few 
particular climate applications, such as probabilistic weather 
prediction~\cite{cano} or causal teleconnection 
analysis~\cite{ebert-uphoff_causal_2012,ebertuphoff_new_2012}.

In recent years, the most popular approach to modeling and obtain the patterns of 
teleconnections in complex systems, like for instance climate,
is based on correlation networks (CNs)~\cite{tsonis_what_2006,Donges2009,donges_backbone_2009,
	boers_prediction_2014,boers_south_2014,boers_complex_2019,zerenner_gaussian_2014,agarwal_network-based_2019}.  
The difference between CNs and BNs is that, whereas the former 
are exclusively built considering pairwise dependencies ({\it e.g.}, correlations) 
between variables (based on the choice of an arbitrary threshold), 
the latter use more sophisticated learning methods 
to model also conditional dependencies, {\it i.e.}, dependencies between two 
(sets of) variables, given a third (set). Here we show that this results in 
sparser, non-redundant, networks with a complex topology, including a 
maximal entropy community structure. Moreover, as we shall show here, from a 
probabilistic perspective CNs and BNs lead to very different 
empirical Gaussian models. The resulting 
CN distribution function is either too simple (mostly dominated by local information, 
therefore, unable to predict teleconnection patterns for high correlation 
threshold values) or too noisy (containing too many parameters for small thresholds) 
and prone to overfitting. In contrast, the 
BN distribution function is a parsimonious representation suitable 
for analysis and probabilistic inference.   

Summarizing, we shall advocate here the use of BNs as non-redundant graph 
representations of complex data, suitable for probabilistic modeling 
and analysis with complex network measures. For the case of climate data~\cite{scutari_who_2018}, 
we shed light on the construction of the proposed BNs and compare with 
the usual CNs approach by characterizing the topology of both type of networks. 
This graph theoretic analysis shows why CN topologies are inherently 
redundant, while BNs are not. Moreover, by using  
machine learning techniques, we analyse the networks as Probabilistic 
Graphical Models (PGMs) that have objective information content 
to soundly present the underlying 
Gaussian model. It will become clear that redundancy in topology is associated 
with the surge of meagre model parameters, so that this redundancy hinders the use
of CNs to extrapolate meaningful features in the case of new data. 
We illustrate this with a particular extrapolation study in which we query the 
networks to predict the teleconnections that appear during a climatic event (El Ni\~no).

\section*{Results}
\subsection*{Climate network construction}

The construction of CNs and BNs in climate is illustrated using a global temperature dataset 
previously used in many studies~\cite{Donges2009, donges_backbone_2009,zerenner_gaussian_2014} 
that shows well-known properties characterized by the interplay 
of strong local and weak long-distant (teleconnections) dependencies. In 
particular, we use monthly surface temperature values on a global $10^\circ$ 
resolution (approx. 1000 km) regular grid for a representative climatic period 
(1981 to 2010), as provided by the ERA interim reanalysis dataset~\cite{dee_erainterim_2011}.
The monthly temperature anomaly values, ({\it i.e.}, the local differences with respect to 
the mean temperature), $X_i$ at grid point $i$ are the 
variables of interest and represent the ($18 \times 36 = 648$) nodes of 
the network models ---the anomaly is obtained by removing the annual cycle 
(the 30-year mean values, month by month) from the raw data. The 
network size (number of edges) and topology of these connections determine the complexity 
and properties of the probabilistic models constructed from the dataset 
and have implications for both model interpretation and 
ability of generalization to new data. 
Hence, we shall discuss the different results in this paper in light of the network size.

\begin{figure}[ht]
	\begin{center}
		\includegraphics[width=\textwidth]{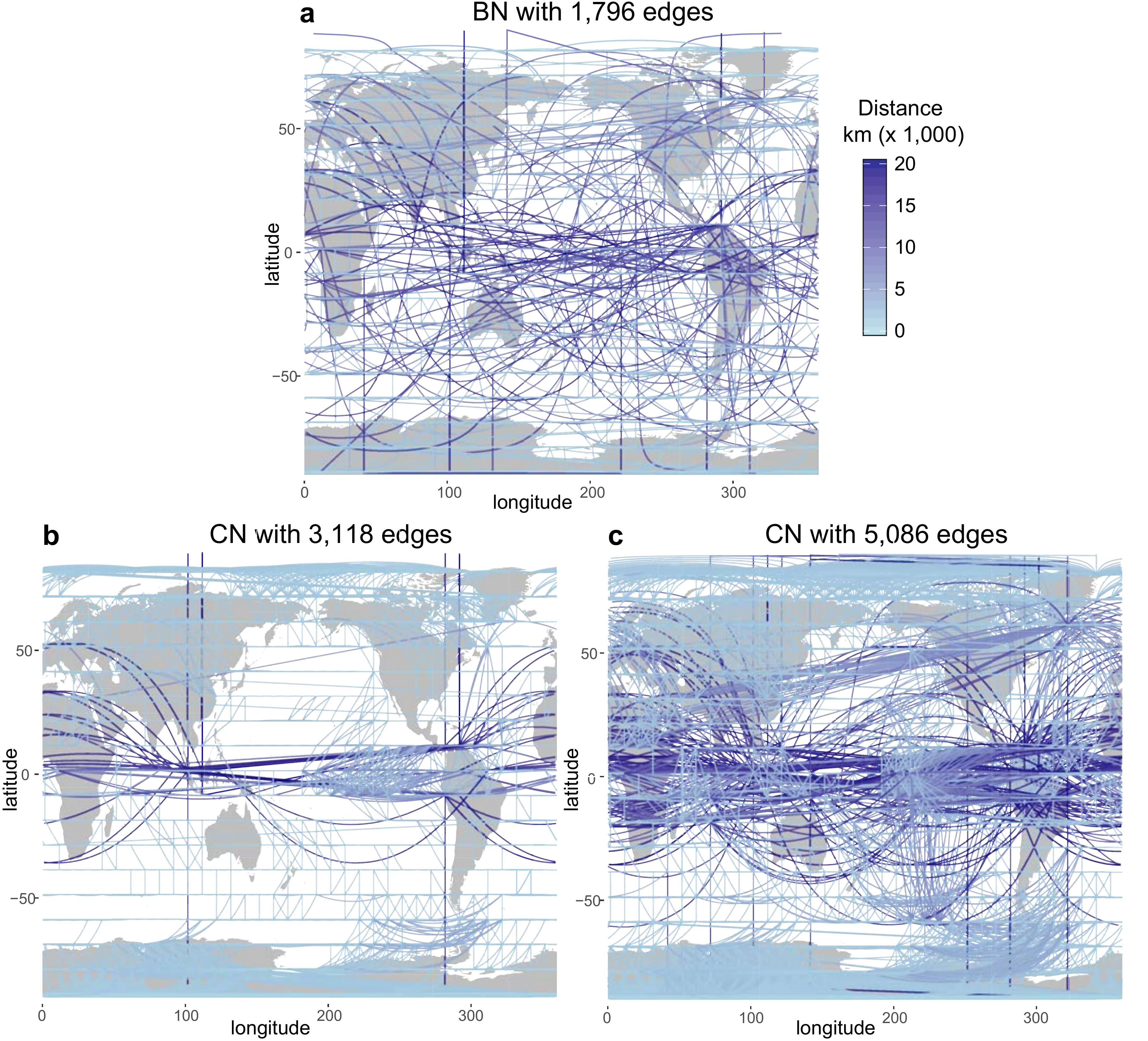}
		\caption{(\textbf{a}) Bayesian Network (BN) with 1,796 edges and Correlation Networks (CNs) consisting of (\textbf{b}) 3,118 edges and (\textbf{c}) 
			5,086 edges. The networks are constructed from 
			monthly surface temperature values on a global $10^\circ$ resolution 
			(approx. 1000 km) regular grid for the period 1981 to 2010. The network 
			represents the dependencies between temperature values in the gridboxes. 
			Edges are coloured as a function of the distance between the gridpoints 
			they connect.}
		\label{fig:networks}
	\end{center}
\end{figure}

The construction of CNs is somehow arbitrary since it is controlled by the 
choice of the threshold on the sample correlation matrix, above which variables
are considered to be {\it connected} in the resulting graph. 
A number of studies have analysed the effects
of different thresholds in the resulting topological properties 
of the network~\cite{tsonis_architecture_2004}. It was found that different 
features of the system are revealed at different 
threshold levels~\cite{donges_backbone_2009} and, as a consequence, the choice of 
the threshold has to reflect a trade-off between the statistical significance 
of connections and the richness of network structures unveiled. 
Small correlation thresholds are needed to capture 
the `weak' teleconnections in the case of climate networks~\cite{Donges2009}, 
but this inevitably leads to a high degree of spurious over-representation of 
the local (strongly correlated) structures, {\it i.e.}, redundancy. 
For example, Figure~\ref{fig:networks}b and~c
show two different CNs 
obtained from the temperature data considering two different illustrative 
thresholds $\tau=0.50$ and $\tau = 0.41$, respectively, 
that yield networks of $3,118$ and $5,086$ 
links. On the one hand, the $\tau=0.50$ CN in 
Figure~\ref{fig:networks}b shows very highly connected 
local regions ({\it e.g.}, the tropics and Antarctica),  and only a few long-distance 
links corresponding to teleconnections. On the other hand, 
the $\tau = 0.41$ CN in Figure~\ref{fig:networks}c
shows a high density of both local and distant links, therefore, a high degree of 
redundancy for characterizing the main physical features. In reality, it 
is difficult to find the {\it appropriate} threshold, or any objective criteria
to select it, to obtain a network that is able to represent the main features 
underlying the data without arbitrariness.

In contrast, BNs are built from data using a machine learning algorithm 
which encodes in a network structure the (marginal and conditional) 
dependencies among the variables that allow to best explain the data 
in probabilistic terms. In this case, the network has a corresponding 
probabilistic model (a Gaussian distribution in this example), given by 
a network-encoded factorization which implies the same underlying dependency 
structure (see the Methods subsection Probabilistic 
BN Models). Learning proceeds iteratively, including new edges 
(dependencies) at each step, so that one maximizes the model likelihood, 
while penalizing complexity (see the Methods section Learning BNs). 
For instance, Figure~\ref{fig:networks}a shows a BN learnt from the 
temperature data with only half of the links as the CN in 
Figure~\ref{fig:networks}b. In contrast to both CNs shown, the BN 
is able to capture both local and long-distant structures 
without redundancy, exhibiting a good balance between local and long distance links.

\subsection*{Community structure}
\begin{figure}[h!]
	\begin{center}
		\includegraphics[width=\textwidth]{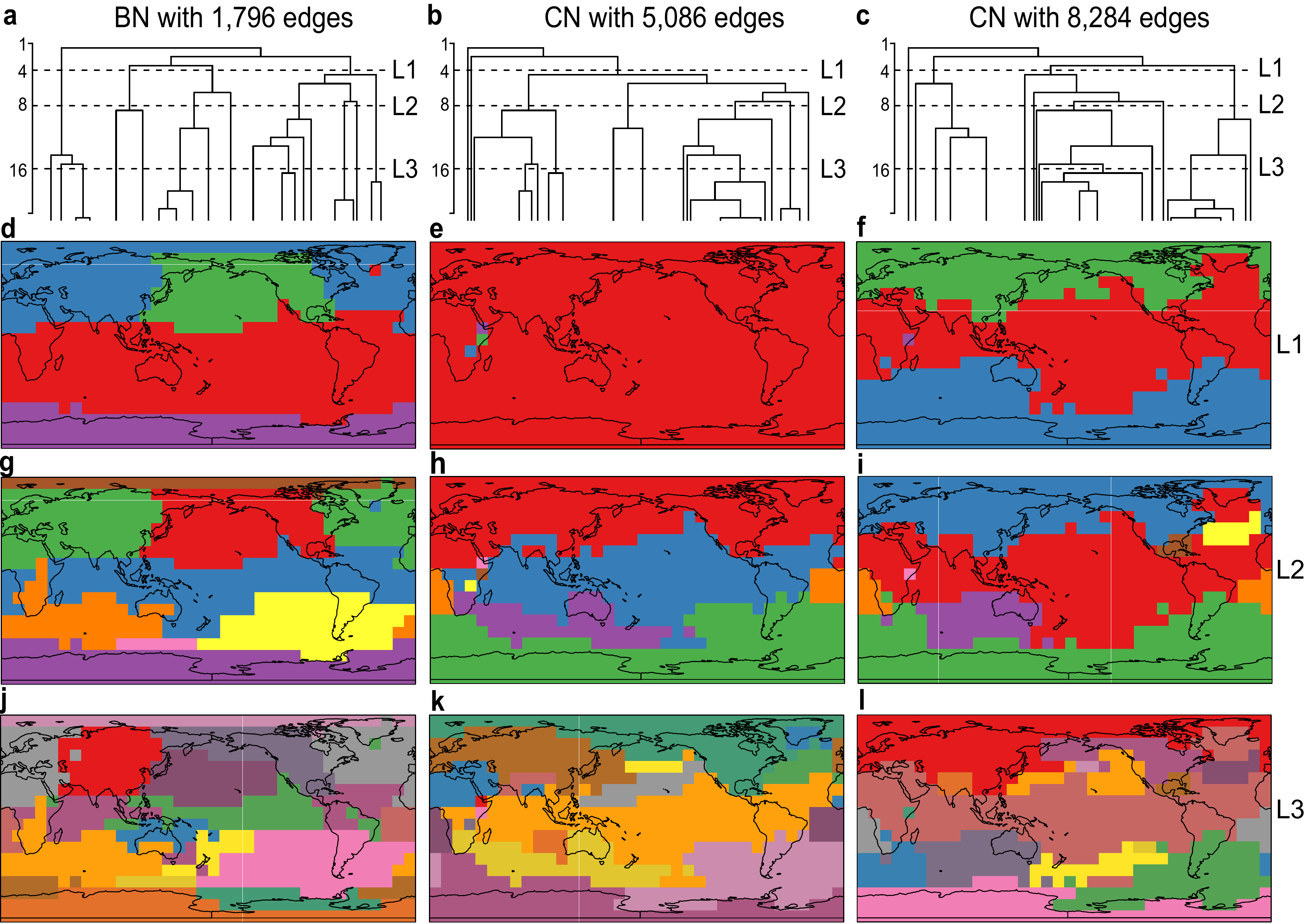}
		\caption{Dendrograms (first row) and community division (2nd to 4th row) 
			of BN with 1,796 edges (first column) and CNs with 5,086 (second column) 
			and 8,284 (third column) edges as found by the edge-betweenness-community 
			detection algorithm. The vertical branches represent communities, which branch 
			off as the algorithm proceeds. The horizontal 
			distance between the two branches adjacent to a
			given branch is an upper bound of the size of that community.
			The height of levels L1, L2 and L3 in the 
			dendrograms indicates the number of communities in which the networks 
			are divided in the 2nd, 3rd and 4th row,
			respectively. Instead of all 648 (number of vertices) levels (divisions 
			of the network by the algorithm) only the first 20 levels are represented 
			in the dendrograms. }
			\label{fig:communities}
	\end{center}
\end{figure}
We deepen the investigation of the 
topology of BNs and CNs of different sizes using well-established 
complex network tools. Results on a selection of complex 
measures that characterize the global and local structure of the networks 
are shown in the Supplementary Material.
Here we focus on the distinctively different community structure 
of BNs and CNs. We analysed the partition of our networks in betweenness-based 
communities. In climate sciences (node or edge) betweenness is an important
proxy that is used to characterize climate topology~\cite{Donges2009,donges_backbone_2009,runge_identifying_2015}. 
The betweenness aims to 
reveal the extent to which edges or nodes are a key for efficient (shortest 
paths) interconnection over distant places on the network. 
Results on a climate data adapted betweenness measure to our BNs and CNs are found in 
Supplementary Material, Figure ~S1.

Climate communities are visually easy to interpret; vertices 
in the same climate community communicate whatever deviation of their mean 
temperature, and the community search algorithm iteratively divides the 
network in a different number of communities allowing the user to visualize 
different scales or levels in the network topology that capture 
different physical features of a network. 
The concepts of communities and betweenness are related. Edges that 
lie between communities can be expected to have high value of 
edge-betweenness (see Methods sections Betweenness centrality and Community detection), 
as such, iterative removal of edges with high betweenness consistently 
splits a network in communities; this technique is used in the 
community search algorithm~\cite{newman_finding_2004,newman_networks:_2010} that
we used to partition our networks in edge-betweenness communities.  

Figure~\ref{fig:communities} shows results on communities for a BN with 
$1,796$ edges and two CNs with $5,086$ ($\tau = 0.41$) 
and $8,284$ ($\tau = 0.33$) edges at three different levels of community partition. 
The BN shows, already at the first partition level, 
Figure~\ref{fig:communities}d, 
a high connectivity among variables in the tropics, the poles and north 
pacific ocean are highlighted. At the second level, Figure~\ref{fig:communities}g, 
the BN  exhibits teleconnections among north and mid Atlantic, east and west 
Pacific, and Indian oceans. Communities continue to 
split as one goes on removing edges with the highest betweenness. 
At the third level, Figure~\ref{fig:communities}j, 
some of the existing communities consists of spatially separated clusters 
that are linked through long-range edges, emphasizing the existence of teleconnections
and its important role in the community structure of BNs.

In contrast, the community partition of the CN of 
size $5,086$ is less informative at the first level, 
Figure~\ref{fig:communities}e. The 
whole globe is fully connected with the exception of three separated gridboxes. 
These three communities correspond to three isolated variables in the 
topology to which the algorithm was forced to assign three uninformative 
communities. In general, the communities arising from CNs contain
less significant information as compared with BNs at the same level
of community partition. This is due to the poor performance of CNs 
in the job of connecting long-distant variables (see 
Supplementary Material Figure~S2 and Figure~S3, in which the size of the largest connected 
component is visualized with respect to the number of edges in the network). As such, 
for all CNs that contain less than 5,086 edges a similar first level is obtained. At the 
second level, Figure~\ref{fig:communities}h, using more communities,
the CN partition captures the connectivity of the tropics, that already 
appears in the BN at the first level, Figure~\ref{fig:communities}d. 
Thereby, some of the separate communities also found at the second BN level, 
Figure~\ref{fig:communities}g, get highlighted. Note that 
the teleconnection of the tropics with the north Atlantic Ocean is not
seen at this early stage of the partition in communities for the CN. 
At the third level, Figure~\ref{fig:communities}k,
many small climate communities pop up in the CN but the presence of very many 
redundant links in the giant 
tropical component, which are also clearly apparent in Figure~\ref{fig:networks}c, 
hinder the algorithm from an efficient partition of the giant component. 
At the third level of the community partitioning of a CN with more edges, Figure~\ref{fig:communities}l ($8,284$ edges), 
the giant tropical community of the CN still remains unbroken. At a deeper level (not shown) 
the giant teleconnected component will be broken by the algorithm after proliferation of 
many redundant communities with little information content.  

The dendrograms in Figure~\ref{fig:communities}a-\ref{fig:communities}c
serve as overviews of the community partition process for the three networks 
discussed above. A significant difference 
in the community fragmentation is apparent: While CNs undergo a 
strongly inhomogeneous division in communities, the BN partitions in a highly uniform fashion.

\begin{figure}[h!]
	\begin{center}
		\includegraphics[width=0.6\textwidth]{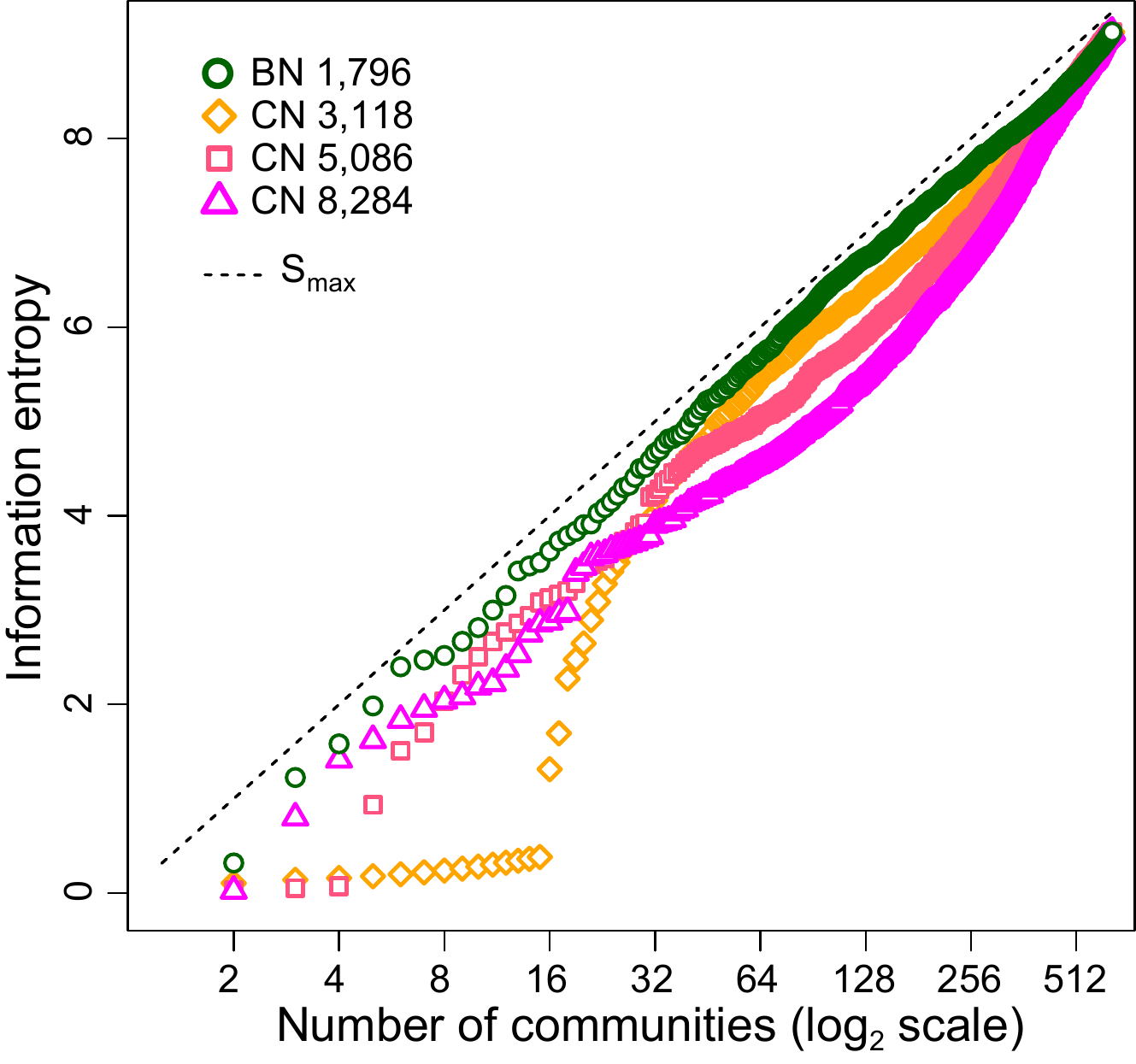}
		\caption{Information entropy $S = - \sum_{\alpha=1}^{N_c}\omega_\alpha \log_2\omega_\alpha$ versus number of disjoint communities $N_c$ in which the network is partitioned. Results are displayed for a BN of 1,796 edges (green) and for CNs of respectively 3,118, 5,086 and 8,284 edges (orange - magenta). The dashed line represents the maximum information entropy $S_{\text{max}}$ that can be obtained for the corresponding number of disjoint communities $N_c$.}
		\label{fig:entropy}
	\end{center}
\end{figure}

These observations can be made quantitative by calculating the 
{\em information entropy}, $S$, of the 
community partition for each type of network (see Methods section Information entropy). 
Suppose that we have our network partitioned in a number, $N_c$, of disjoint 
communities and we ask ourselves what is the information content of this 
community structure. In other words, how much information gain (on average) 
we would obtain by determining that a random node belongs to a certain 
community. If the entropy is high, this means that every time we ascertain 
that a site belongs to a given community we gain much information 
on the structure. Conversely, if entropy is low the average 
information gain we obtain by this process is 
small, on average. The maximum entropy corresponds to an even distribution of the sites among 
existing communities, while a low entropy would mean that some communities are much 
larger than others, and so, there is a much higher probability that a 
site, picked at random, belongs to the most populated communities. 
The amount of information conveyed in this case by 
specifying the community structure is lower. One can prove (see Methods 
section Information entropy) that the maximum entropy
corresponds to $S_\mathrm{max} = \log_2 N_c$, where $N_c$ is the number of communities.   
In Figure~\ref{fig:entropy} we plot $S$ as a function of the 
number of communities for the optimal BN and 
CNs of different sizes. One can see that $S$ for the optimal BN 
of $1,796$ edges, which corresponds 
to that in Figure~\ref{fig:networks}a, attains values 
close to the maximal information entropy for {\em any} number of
communities $N_c$, from early stages in the community splitting process 
(where only $4$ to $6$ communities are present) to later stages 
(when a few hundred communities have been found).
In contrast, the information entropy of CNs (no matter the threshold chosen) 
is always below the BN optimal case. This clearly shows that the community 
structure of small, sparse BNs have much larger amounts of information 
content than the CN counterparts.

\subsection*{Probabilistic model construction and cross-validation}
\begin{figure}
	\begin{center}
		\includegraphics[width=\textwidth]{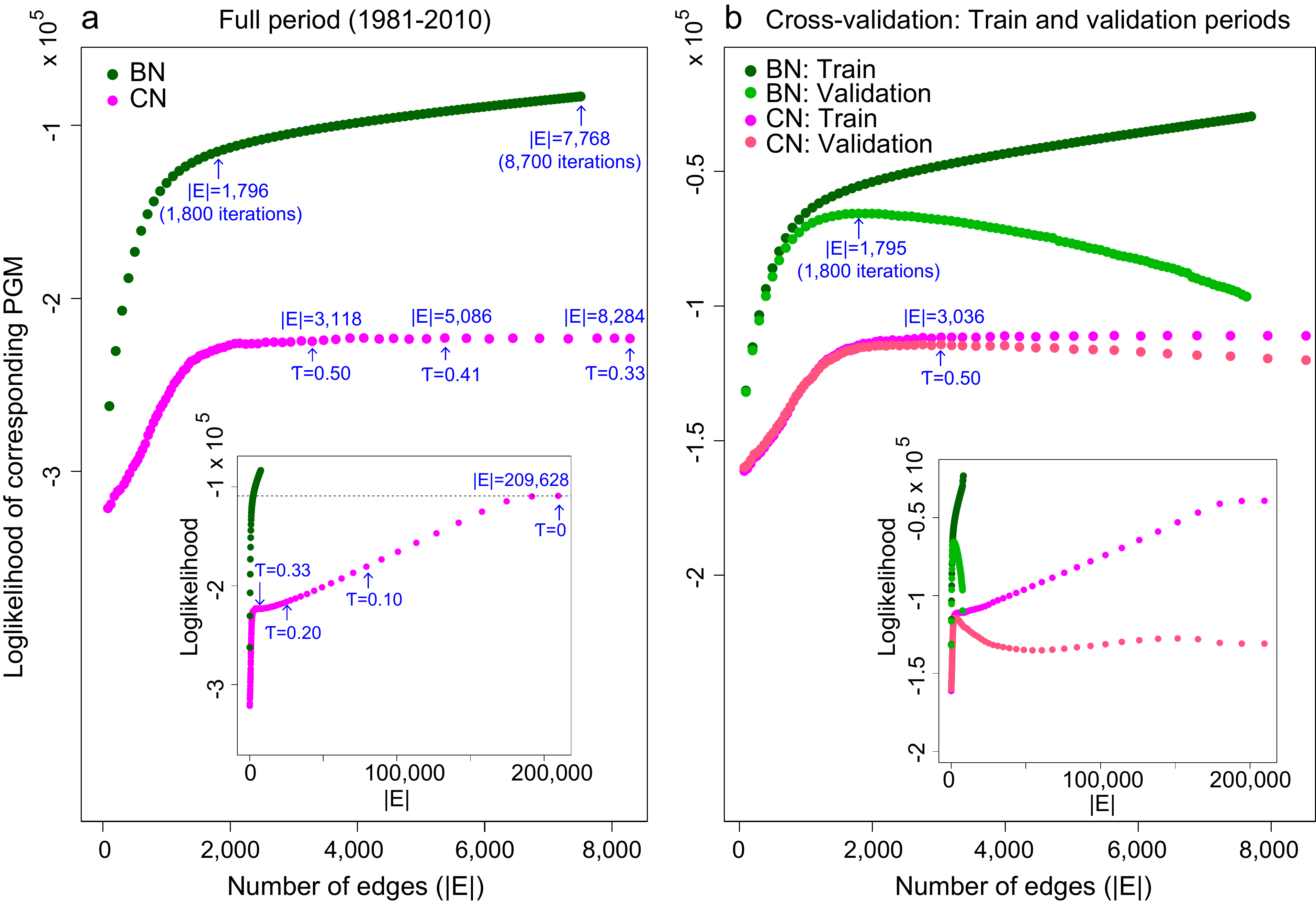}
		\caption{(\textbf{a}) Log-likelihood values  
			$\log\Prob(\D_\mathrm{c}\given PGM_\mathrm{c})$ versus 
			number of edges ($|E|$) of Bayesian (green) and correlation (magenta) PGMs 
			that are learnt from the complete dataset $\D_\mathrm{c}$ 
			and (\textbf{b}) Log-likelihood 
			values $\log\Prob(\D_\mathrm{t}\given PGM_\mathrm{t})$ (green and magenta) 
			and 
			$\log\Prob(\D_\mathrm{v}\given PGM_\mathrm{t})$ (light green and light 
			magenta) versus number of edges of Bayesian (green and light green) and 
			correlation (magenta and light magenta) PGMs 
			that are learnt with the train dataset $\D_{t}$. In (\textbf{a}) and (\textbf{b}) 
			outer windows are a zoom of the small windows inside. Some log-likelihood 
			values of correlation PGMs are labeled with the threshold $\tau$ 
			which was used to construct 
			the CN and some log-likelihood values of Bayesian PGMs are labeled with the number of iterations that was used by the structure learning algorithm to construct the BN. In (\textbf{b}) these labels are placed by the BN and CN for which the corresponding PGM obtains a maximum value of $\log\Prob(\D_\mathrm{v}\given PGM_\mathrm{t})$. 
			In the small window of (\textbf{a}) the dotted line
			represents the log-likelihood value of a complete Correlation PGM 
			of size 209,628, corresponding to $\tau = 0$.}	
			\label{fig:logliks}	
	\end{center}
\end{figure}
Here we analyse from a probabilistic perspective the networks built in the previous sections 
by extending the graphs to full Probabilistic Graphical Models (PGMs), in which edges on the graph will represent parameters 
in the probability density function.  In this paper we work with the
natural choice of multivariate 
Gaussian distributions as PGMs. On the one hand, the probability density  
function for a CN is constructed by specifying the covariance matrix 
elements $\Sigma_{ij}$ from the empirical correlations that are above 
the fixed threshold ({\it i.e.}, for those edges that are 
present in the CN graph), and $\Sigma_{ij} = 0$ otherwise. 
On the other hand, the probability distribution function 
for a BN is represented by a factorization of conditional multivariate 
Gaussian probabilities and the parameters are the linear regression 
coefficients of a variable on its conditioning variables ({\it i.e.}, 
those that are connected by an edge in the graph with a parent-child relation). 
See the methods section Probabilistic 
Gaussian Graphical Models for more information on the extension of CNs and 
BNs into PGMs and their particular encoding of the multivariate Gaussian 
density function.

A key problem in machine learning is whether the models learnt from a 
training data sample can capture general and robust features of the 
underlying problem, thus providing out-of-sample extrapolation capabilities. 
This property is known as {\em generalization} and it is typically assessed 
in practice using a test data sample (or, more generally, by cross-validation) 
to check whether the model is {\em overfitted} (the model explains very well 
the training data but fails to explain the test). 

Once a network is extended to a PGM one can measure the goodness of fit of 
that PGM to any dataset $\cal D$ using the log-likelihood 
$\log P({\cal D}|\mathrm{PGM})$. This
quantity can be interpreted as the probability of a given dataset $\cal D$ 
when $P$ is modelled by a certain PGM (see Methods Section 
Log-likelihood definition and calculation, for details). 
The log-likelihood compares models that encode the same type of 
density function $P$, but with different parameters, and 
should be interpreted comparatively; the
log-likelihood value of model A is not very meaningful in absolute 
terms, however, if log-likelihood of model A is higher than that of 
model B, one can conclude that model A explains the data 
better than model B. In this work all types of PGMs and, along 
them, all networks of different sizes
encode a multivariate Gaussian distribution over a constant 
dimensional variable space, making the log-likelihood an adequate 
comparative measure~\cite{Koller:2009:PGM:1795555}.

First, we use the log-likelihood to measure the goodness of fit of CNs 
and BNs by calculating $P({\cal D}_\mathrm{c}|\mathrm{PGM_{c}})$ for networks of various sizes, 
in which $\mathrm{PGM_{c}}$ refers to the PGM that is learnt from the complete 
dataset ${\cal D}_\mathrm{c}$. Next, we use the log-likelihood to assess the 
{\em generalization} capability of the models, calculating 
cross-validated log-likelihood values $P({\cal D}_\mathrm{v}|\mathrm{PGM_{t}})$, obtained by 
splitting the data in two halfs, one for training ${\cal D}_\mathrm{t}$ 
and one for validation ${\cal D}_\mathrm{v}$, where $\mathrm{PGM_{t}}$ denotes the PGM that is 
learnt from from the training dataset ${\cal D}_\mathrm{t}$. 
Figure~\ref{fig:logliks}a and \ref{fig:logliks}b show the results for the 
goodness of fit and generalization, respectively.

Figure~\ref{fig:logliks}a shows 
$P({\cal D}_\mathrm{c}|\mathrm{PGM_{c}})$ as a function of the 
network size for CNs and BNs. Addition of parameters to a model facilitates 
it to explain the data on which it was trained and, thus, this should increase 
log-likelihood. Figure~\ref{fig:logliks}a shows that the amount of gain
in log-likelihood depends on the type of model; adding parameters (edges) 
to the BN turns out to be very efficient, yielding a gain in 
log-likelihood. However, when adding parameters to the 
CN it becomes efficient only up to a certain size (around $2\times10^3$ edges). Once 
this size is exceeded the log-likelihood only continues to grow after a 
great amount of parameters have been added (indeed, the growth continues around 
$3\times10^4$ edges, see inset of Figure~\ref{fig:logliks}a).
The figure shows that CNs and BNs of similar size strongly differ in the 
amount of data their associated PGMs explain, BNs being much more efficient 
in explaining the data. The plateau observed in the log-likelihood curve 
for the CN model indicates the existence of a range of correlations that 
mostly represent redundant parameters. 

Figure~\ref{fig:logliks}b shows the training, 
$P({\cal D}_\mathrm{t}|\mathrm{PGM_{t}})$, and
validation, $P({\cal D}_\mathrm{v}|\mathrm{PGM_{t}})$, log-likelihoods for 
CNs and BNs. As one can see in the plot, the log-likelihood of 
$P({\cal D}_\mathrm{t}|\mathrm{PGM_{t}})$ in Figure~\ref{fig:logliks}b
is consistent with that of $P({\cal D}_\mathrm{c}|\mathrm{PGM_{c}})$
(in Figure~\ref{fig:logliks}a) for both BN and CN, 
showing that the PGMs learnt from the (halved) training sets ${\cal D}_\mathrm{t}$ 
are as good as those obtained with the complete set of data, 
${\cal D}_\mathrm{c}$, in both type of networks. 
As for the validation, the log-likelihood of $P({\cal D}_\mathrm{v}|\mathrm{PGM_{t}})$ shows 
that both CNs and BNs exhibit an `optimal' size for which the excluded 
validation data is explained best. PGMs with a number of edges 
(parameters) above the optimum are overfitting the data that were used to 
train the models and fail to generalize out-of-sample (validation) 
datasets. The log-likelihood curve of the BN model declines 
after the maximum, located around $1,795$ edges. Indicating that the PGM 
is performing worse as we include more edges. Similarly, 
the log-likelihood curve of the CNs declines after a maximum 
at $3,118$ edges. Note that, for CNs, in the range of 
network sizes where Figure~\ref{fig:logliks}a showed a plateau, the 
validation log-likelihood declines dramatically. Therefore, CNs with a 
correlation threshold above $\tau = 0.56$ result in a generalizable PGM. 
Edges/parameters for $\tau$ between $0.20$ and $0.56$ still represent 
relatively strong correlations but these CNs are not generalizable to explain new data. 
The test log-likelihood  curve (inset of Figure~\ref{fig:logliks}b) 
has a small revival when edges with correlation smaller than 
$\tau = 0.2$ are added. One may conclude that, in CNs, relatively strong correlations 
are not always relevant and small correlations are not always negligible
when constructing the corresponding PGM. This is 
due to the mixing of strong but short and weak but relevant long-range spatial 
correlations, which significance CNs cannot capture effectively. 
Placing links/parameters by a CNs approach 
easily leads to overfitting of high correlations and underestimation of 
the effect of small (but physically important, teleconnections in 
this case) dependencies.

\subsection*{Estimating Conditional Probabilities}
The estimation of conditional probabilities is one of the key problems 
in machine learning and a number of methodologies have been proposed for 
this task, such as regression trees~\cite{Beygelzimer:2009} or Support 
Vector Machines~\cite{856437}. Multivariate Gaussian distributions provide 
a straightforward approach to this problem allowing to estimate the impact 
of an evidential variable $X_e$ (with known value) to other variables 
(gridboxes in this study). For example, assuming warming conditions in a 
particular gridbox of the globe $X_e$ (e.g. a strong increase in 
temperature, say $X_e = 2\sigma_{X_e}$) the conditional probability of 
the other gridboxes $P(X_i|X_e)$ provides a quantification of physical 
impact of this evidence in nearby or distant regions. This will allow, 
for instance, to study teleconnections of $X_e$ with other distant regions.

\begin{figure}[h!]
	\begin{center}
		\includegraphics[width=\textwidth]{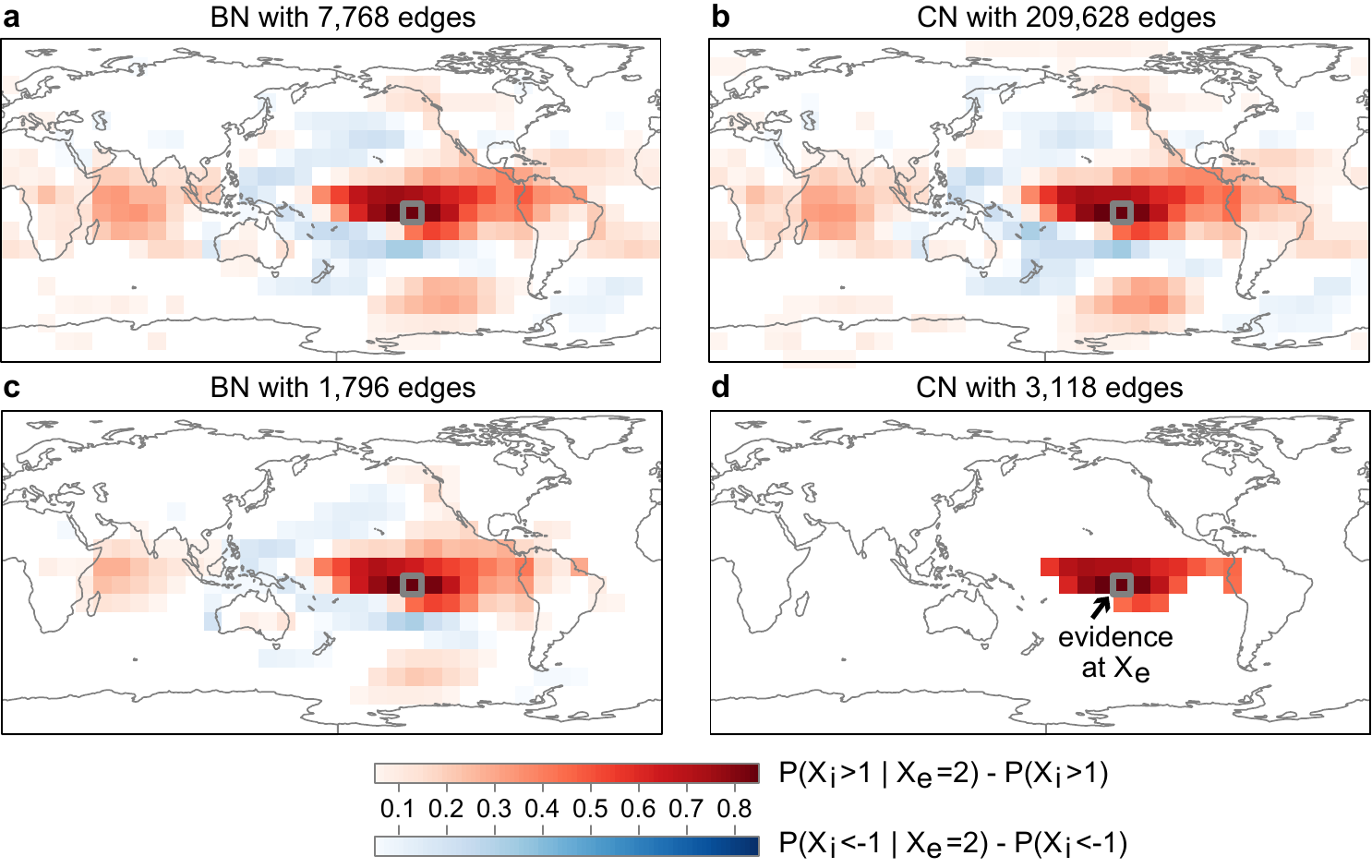}
		\caption{Differences of conditional and marginal probabilities 
			$\Prob(X_i \geq 1\given X_e = 2) - 
			\Prob(X_i \geq 1)$ (red scale) and  
			$\Prob(X_i \leq 1\given X_e = 2) - 
			\Prob(X_i \leq 1)$ (blue scale) as modelled with Bayesian 
			PGMs with (\textbf{a}) 7,768 and (\textbf{c}) 1,796 edges and 
			correlation PGMs with (\textbf{b}) 209,628 and (\textbf{d}) 3,118 edges. The location 
			of the evidence variable $X_e$ 
			is denoted with a grey box and labeled in (\textbf{d}) with ``Evidence at $X_e$". The event $X_e = 2$ indicates a positive deviation of the mean value of twice its standard deviation, i.e. strong warming in $X_e$. 
			The Bayesian (\textbf{a}) and Correlation PGM (\textbf{b}) in 
			the first row are heavily parametrized. The Bayesian (\textbf{c}) and Correlation 
			PGM (\textbf{d}) in the second row correspond to cross-validated optimal PGMs.}
			\label{fig:evidence}	
	\end{center}
\end{figure}

We illustrate the performance of CN and BN PGMs to estimate conditional 
probabilities with the case study of the east Pacific ocean 
teleconnections--- El Ni\~no - Southern Oscillation 
teleconnections~\cite{kajtar_tropical_2017}---,  selecting a particular 
gridpoint $X_e$ in the equatorial pacific (grey box in Figure~\ref{fig:evidence}a-\ref{fig:evidence}d). 
A single map in Figure~\ref{fig:evidence} visualizes the conditional probabilities of warming 
and cooling conditions for all 
other gridboxes $X_i$ ({\it i.e.}, $P(X_i \geq \sigma_{X_i}|X_e = 2\sigma_{X_e})$ and 
$P(X_i \leq \sigma_{X_i}|X_e = 2\sigma_{X_e})$, respectively) given very 
warm conditions at $X_e$. The four different maps display the results 
of four PGMs, two BNs and two CNs. In Figure~\ref{fig:evidence}a and \ref{fig:evidence}b 
the results for heavily parametrized (heavily overfitted) Bayesian ($7,768$ edges) and correlation ($209,628$ edges) PGMs are shown. These models exhibit similar results, showing warm 
deviation in teleconnected regions in the Indic and Southern Pacific 
and Atlantic oceans. The existence of these teleconnections is in agreement with the 
literature~\cite{kajtar_tropical_2017}. Figure~\ref{fig:evidence}c-\ref{fig:evidence}d
show the probabilities for the networks with optimum size
according to our cross-validation tests ({\it i.e.}, Figure~\ref{fig:evidence}c 
a BN with $1,796$ edges and Figure~\ref{fig:evidence}d a 
CN with $3,118$ edges). The CN in panel Figure~\ref{fig:evidence}d only 
captures local deviations (note the low 
value of $P(\D_{c}|\mathrm{PGM_{c}})$ in Figure~\ref{fig:logliks}a). 
Teleconnections are only quantitatively 
captured in CNs of greater size with higher log-likelihood 
$P(\D_{c}|\mathrm{PGM_{c}})$, 
however, as shown in the above section, these models are highly overfitted and, 
therefore, lack of generalization capabilities ({\it i.e.} they can only 
explain the data used for training and do not posses generalizable 
physical relationships--- generalizable teleconnections in this 
case---). Drawing conclusions on the strength of the captured 
(teleconnected) dependencies and making decisions on the basis of 
predictions in CNs of large sizes is therefore questionable, if not 
plain wrong, as shown by our cross-validation tests above. 
On the other hand, the cross-validated optimum BN with $1,796$ 
edges in Figure~\ref{fig:evidence}c does capture 
the teleconnections (with smaller probability in some cases) and
$P(\D_{v}|\mathrm{PGM_{t}})$ gives higher values as compared with 
those in the CNs, therefore, it is generalizable to explain new data. 
The reason why probabilities are a bit smaller for some teleconnected 
regions (with respect to the heavily parametrized model in 
Figure~\ref{fig:evidence}a) is the 
non-stationary nature of El Ni\~no events, which can take various forms, 
{\it e.g.}, the Cold Tong Ni\~no event~ and the  Warm Pool 
(Modoki\cite{ashok_nino_2007}) Ni\~no event\cite{kug_two_2009}, the former exhibiting 
stronger surface temperature teleconnection with the Indian Dipole 
and the latter with the teleconnected regions at higher 
latitudes \cite{kug_two_2009,dimri_warm_2017,hu_analysis_2012,jadhav_possible_2015,sun_impacts_2013}. 
Low but non-zero probability on significant deviation in teleconnected 
regions is, thus, a truthful presentation of the impact of the evidence if 
this is to be generalized to different El Ni\~no types co-existing in 
the dataset.

 

\section*{Discussion}

Networks are the main subject of study in complex network theory, whereas from a machine learning perspective networks has been supporting tools to obtain probabilistic models.
In this work we show that BNs, developed by the machine learning community, constitute an extremely appealing and sound approach to build complex data-driven networks based on a probabilistic framework. BNs give an optimal, non-redundant, probabilistic Gaussian model of the complex system of interest using a network support characterizing the relevant dependencies. The resulting networks are sparse but rich in topological information as shown by standard complex network measures, while the probabilistic counterparts are sound models generalizable to new data and, therefore, have predictive power. In contrast, we have shown that the most common approach to graphically model complex systems, based on CNs constructed from pairwise correlations, is prone to overfitting, depends on an arbitrary threshold, and performs very poorly when one intends to generalize to explain new data, therefore, lacking of  predictive power. We have shown that sparse networks that have predictive power are particularly useful when studying long distant connections in a complex system. 
We studied the teleconnections that occur in a complex climate dataset and showed that BNs without further post-processing faithfully reveal the various long-range teleconnections relevant in the dataset, in particular those emerging in El Niño periods.

We proposed to find the optimal size (number of edges) of a network simply and generically from the corresponding log-likelihood of the
data-driven PGM. The log-likelihood measures the ability of a PGM to explain 
the data. Log-likelihood plots clearly show that there exists a region where 
the gain in explanatory power by adding more parameters/edges dramatically
slows down, as reflected by a change in the slope 
of the log-likelihood curve in Figure~\ref{fig:logliks}. A probabilistic model, either 
BN or CN based, begins to be overfitted once the log-likelihood curve bends, 
giving an objective, non biased, estimate for the 
optimal number of edges that need to be used. For both BNs and CNs, it turns out that
including more edges results in little gain in log-likelihood and tends to produce 
progressively overfitted models, leading to less capability for explaining new data. 
We have shown that CNs need to go well above this optimum in order to capture 
weak, but important, teleconnections; while BNs capture the significant (even if small)
teleconnections early on, when only a few hundreds of edges have been included in the model. 

In addition, we have shown that the edge-betweenness community structure of BNs 
attains nearly maximal information entropy, $S_\mathrm{max} = \log_2 N_c$, when the number of 
edges is around the optimum, no matter the number of communities $N_c$. In this sense,
the optimum number of parameters (number of edges) in a BN 
is an objective non arbitrary quantity.
From an information theory perspective, this means that each assignation of a 
node to a given community gives maximal information about the community structure,   
reflecting the fact that virtually no edge is redundant. In contrast, the entropy
of the community partition for CNs is far below the maximum, unless several tens 
(or even hundreds, depending on the correlation threshold)
of communities are taken into account.

The choice of the threshold, i.e. determining the amount of edges, in CNs is problematic. Donges {\it et al.}~\cite{Donges2009,donges_backbone_2009} already noted that to 
capture, for example, teleconnections with topological measures the correlation 
threshold has to be chosen below some maximum value. This small threshold does not coincide with that needed for the network to be statistically most significant. In \cite{Donges2009} various thresholds are chosen in function of  topological measures to ensure a balance between the structural richness that is unveiled by the measure and the statistical significance of the network. Usually, once the threshold is chosen, this choice is justified by conducting a robustness analysis testing the effect of the threshold on the qualitative 
results and/or assuring a minimum level of significance--- using 
significance tests based on randomly shuffled time series, Fourier 
surrogates and twin surrogates--- ~\cite{tsonis_architecture_2004,Donges2009,donges_backbone_2009}. This approach poses several problems for the practical construction and 
interpretation of these models. Recent studies, mostly in the context of extreme rainfall data, thus
propose to include an 
extra {\it ad-hoc} post-processing step in the network construction phase, in 
which insignificant edges (probable to occur in a random network) are removed from the final network in order to alleviate the problems introduced by redundancy~\cite{boers_complex_2019,boers_extreme_2014,boers_prediction_2014,boers_south_2014}; Boers {\it et al.}~\cite{boers_complex_2019} even correct their extreme rainfall network data two times, firstly, by keeping only significant links with respect to a random network and 
later removing links that are not part of a `link bundle', 
{\it i.e.}, not `confirmed by other links'. As shown by the present work, there is a 
fundamental difference between the construction of CNs and that of 
BNs. On the one hand, CNs capture `strong relationships' early on in the construction process
and are affected by troublesome overfitting problems that would eventually 
need to be `cured' by some some type of post-processing to maintain only 
the statistical significant ones among them--- a job for which no general 
unbiased solution exists. On the other hand, the BN construction we propose
here only captures statistically significant relationships (no matter if weak or strong) 
and reveals which of them are essential for increasing the explanatory capability 
of the model (evidence propagation).

To avoid erroneous generalization of the relative strength of significant connections in complex networks we advocate in this paper for the use of BNs, which generically 
yield sparse, non redundant, maximal information containing, and generalizable 
networks suitable for extracting qualitative information with complex measures, but that also explain new data and do not require any {\it ad-hoc} extra correction steps. 

\section*{Methods}
\subsection*{Learning BN structure from data}
A BN is estimated  with the help of a structure learning 
algorithm that finds the conditional dependencies between the 
variables and encodes this information in a Directed Acyclic Graph (DAG). 
Graphical (dis-)connection in the DAG implies conditional 
(in-)dependence in probability. From the structure of a BN a factorization 
of the underlying joint probability function P of the multivariate random 
variable $\X$ (as given by equation~(\ref{eq:mvg-fac})) can be deduced. We shall
come back to this factorization in the Methods Section Probabilistic 
BN Models where we explain how networks can be extended 
to their corresponding Probabilistic Graphical Models (PGMs). 

In general there are three types of structure learning algorithms: 
constrained-based, score-based, and hybrid structure learning algorithms--- 
the latter being a combination of the first two algorithms.

Constrained-based algorithms use conditional independence tests of the 
form $\mathrm{Test}(X_i,X_j|S;\D)$ with increasingly large candidate separating 
sets $S_{X_i,X_j}$ to decide whether two variables $X_i$ and $X_j$ are 
conditionally independent. All constraint-based algorithms are based 
on the work of Pearl on causal graphical 
models~\cite{Verma:1990:ESC:647233.719736} and its first practical 
implementation was found in the Principal Components 
algorithm~\cite{spirtes_causation_1993}. 
In contrast, score-based algorithms apply general machine learning 
optimization techniques to learn the structure of a BN. Each candidate 
network is assigned a network score reflecting its goodness of fit, 
which the algorithm then attempts to maximise~\cite{russell_artificial_1995}.
Somewhere else some of us~\cite{scutari_who_2018, scutari_who_2019} 
compared structure learning 
algorithms belonging to the three different classes on accuracy and 
speed for the climate dataset used here. The comparison was based on the 
intrinsic edge-attachment method after removing the confounding effect 
of statistical criteria (in the form of scores/independent tests). We 
found that score-based algorithms perform best for the 
complex data in the climate data set. Algorithms in this class are able 
to handle high-variable-low-sample size data and find networks of all 
desired sizes. Constrained-based algorithms only can model complex data 
up to a certain size and, as a consequence, for climate data they only reveal
local network topology. Hybrid algorithms perform better than 
constrained-based algorithms on complex data, but
worse than score-based algorithms.

\begin{figure}
	\begin{center}
		\includegraphics[width=0.6\textwidth]{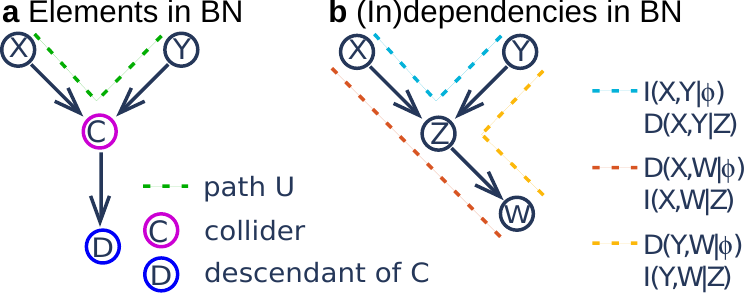}
		\caption{(\textbf{a}) Nomenclature of elements in a Bayesian Network (BN) and (\textbf{b}) Some (in)dependencies in a simple BN consisting 
			of the nodes X,Y,Z and W. Two sets of nodes are dependent given a third if conditions (1) and (2) in the main text are fulfilled. In (\textbf{b}), on the one hand, the conditional 
			relationship $X,Y|Z$ and the marginal relationships 
			$X,W|\emptyset$ and $Y,W|\emptyset$ satisfy conditions (1) and (2), 
			so that we have $D(X,Y|Z)$, $D(X,W|\emptyset)$ and $D(Y,W|\emptyset)$. 
			On the other hand, the marginal relationship $X,Y|\emptyset$ violates 
			condition (1) and the conditional relationships $X,W|Z$ and $Y,W|Z$ 
			violate condition (2), so that we have $I(X,Y|\emptyset)$ and $I(X,W|Z)$ and $I(Y,W|Z)$. \\
			\textbf{Proof of $\bm{D(X,Y|Z)}$ - Conditional dependence of X and Y given Z in Figure~6\textbf{b}.}
			The conditioning set S exists of Z. The only path between X and Y is the blue path. Hence we declare the blue path U. Z is a collider and Z is in S. There are no other colliders on U. Hence condition (1) is satisfied. Z is the only variable on U. And Z is a collider. Thus, U does not contain non-colliders. Hence condition (2) is satisfied. As condition (1) and (2) are satisfied we have that X and Y are dependent given Z, i.e. $D(X,Y|Z)$.}
			\label{fig:dependencies}		
	\end{center}
\end{figure}

In the following we describe how a DAG, found by a structure learning algorithm, encodes
conditional dependencies. New nomenclature is indicated with an asterisk and illustrated in Figure~\ref{fig:dependencies}a.
Two nodes X and Y are conditionally dependent given a set S (denoted by $D(X,Y|S)$) if and only 
if they are graphically connected, that is,
if and only if there exists a path $\text{U}^*$ between X and Y satisfying the following two
conditions: 
\begin{itemize}
	\item {\em Condition (1)}: for 
	every collider$^*$ C (node C such that the part of U that goes over C has the 
	form of a V-structure, {\it i.e.},  $\rightarrow C  \leftarrow$) on U, either 
	C or a descendant$^*$ of C is in S.
	\item {\em Condition (2)}: no non-collider on U is in S. 
\end{itemize}
If the above conditions do not hold we call X and Y conditionally 
independent given the set S (denoted by $I(X,Y|S)$). 
Marginal dependency between two nodes can be encoded 
by any path U with no V-structures.
In Figure~\ref{fig:dependencies}b six conditional (in)dependence statements are highlighted in a simple DAG. In the caption of Figure~\ref{fig:dependencies} one of the statements is proved at the hand of conditions (1) and (2). 

In this work we use a simple score-based algorithm, the Hill Climbing 
(HC) algorithm~\cite{russell_artificial_1995}, to learn BN structure.
The HC algorithm starts with an empty graph and iteratively adds, removes 
or reverses an edge maximizing the score function. We used the Bayesian 
Information Criteria (BIC) (corresponding to $\BIC_0$ in 
\cite{scutari_who_2018,scutari_who_2019}) score, which is defined as: 
\begin{equation}
	\BIC(\G; \D) =
	\sum_{i=1}^N \left[\; \log \Prob(\XPi) - \frac{|\T|}{2}\log N \;\right].
	\label{eq:bic}
\end{equation} 
P refers to the probability density function that can be deduced from 
the graph (see Methods Section Probabilistic BN 
Models.), $\Pi_{X_{i}}$ refer to the parents of $X_i$ in the graph 
({\it i.e.} nodes Y with relation $Y \rightarrow X_i$ in the graph) and 
$|\Theta_{X_{i}}|$ is the amount of parameters of the local 
density function $\Prob(\XPi)$.

\subsection*{Betweenness centrality}
The betweenness centrality measures the extent to which a node lies on 
paths between other nodes\cite{newman_networks:_2010}. A node is assigned high betweenness 
centrality if it is traversed by a large number of all existing shortest 
paths (geodesics). We define $g_{jk} $ as the total number of geodesics 
between node $X_j$ and node $X_k$ and $g^i_{jk}$ as the number of geodesics 
between node $X_j$ and node $X_k$ that include $X_i$.
Then, the betweenness centrality $BC_i$ of node $X_i$ can be expressed as 
\begin{equation}\label{eq:betweenness}
	BC_i = \sum_{j, k \neq i}^{N} \frac{g^i_{jk}}{g_{jk}},
\end{equation}
with the convention that $g^i_{jk}/ g_{jk} = 0$ if 
both $g^i_{jk}$ and $g_{jk}$ are zero.

\subsection*{Community detection}
A community (also group or cluster) is formed by 
sets of nodes that are tightly knit with many edges to 
other nodes inside the set, while there are few edges 
connecting the set with other sets. A transparent way of finding 
communities is to look for edges that lie between communities and 
remove them. In this way one is left with just the isolated communities. 
One can detect `edges between communities' noting that those edges 
typically have high values of edge betweenness centrality. Edge 
betweenness of a given edge is defined\cite{newman_networks:_2010} in a similar matter 
as the node betweenness in equation~(\ref{eq:betweenness}); instead of 
defining $g^i_{jk}$ as the number of geodesic paths that 
run along a node, $g^i_{jk}$ computes 
the number of geodesic paths that run along the edge $i$ and the sum is 
over all nodes $j \neq k$. Based on this definition one expects that 
edges between communities have higher values of edge betweenness with 
respect to those that are not between communities because all geodesics 
between two nodes in two different communities go over the first. 
The betweenness-based-community detection algorithm is then as follows: The 
algorithm\cite{freeman_centrality_1978,brandes_faster_2001} starts with one community that contains all nodes, then 
iteratively splits this giant community in other communities by removing 
edges with the highest edge-betweenness value partitioning 
the network in smaller communities, step by step. This continues 
until all nodes are singleton communities. In the process 
edge betweenness values of edges 
will change because shortest paths are rerouted after an edge removal, 
hence the edge betweenness values are recalculated at every step. 
The splitting process can be represented in a dendrogram, showing the 
division of larger communities into smaller ones at every stage of 
the algorithm evolution.

\subsection*{Information entropy}
In order to quantitatively measure the information content of communities 
we compute the entropy\cite{Shannon_mathematical_1949} of any given community partition. 
At a given level of the partitioning 
process we label the existing communities with $\alpha = 1, \dots, N_c$,
where $N_c$ is the number of communities, and define the entropy of the community partition 
as $S = - \sum_{\alpha=1}^{N_c}\omega_\alpha \log_2\omega_\alpha$, 
where $\omega_\alpha$ is the fraction of nodes that belong to the $\alpha$-community.  
This entropy is a measure of the amount information encoded in the community 
size distribution. If we were to store the complete community list and its members 
by specifying to which community, $\alpha = 1, \dots, N_c$, 
each site, $i = 1, \dots, N$, belongs to then $S$ would tell us the 
average amount of information,
$N S(N_c)$, that would be required to do the job. The entropy is maximal 
when the $N$ network sites are evenly distributed among the available
$N_c$ communities. This corresponds to 
$w_\alpha = \omega$, where $\omega = 1/N_c$ for all $\alpha$, then the entropy becomes 
$S_\mathrm{max} = -N_c \, \omega \log_2\omega = \log_2 N_c$. 
Any entropy below this number means the sizes of 
communities are uneven, more so the lower the entropy. A lower entropy for a community partition
means less information content is stored in the community structure. Entropy is then a 
proxy to measure redundancy.

\subsection*{Probabilistic Gaussian Graphical Models (PGGMs)}
The term refers to the choice of a multivariate 
Gaussian joint probability density (JPD) function to associate graph
edges with model parameters in a given PGM, such that
the probabilistic model encodes in the JPD function a large number of 
random variables that interact in a complex way with each other by a 
graphical model. A graphical model exists from a graph and a set of 
parameters. The set of parameters characterize the JPD function 
and are reflected in the corresponding graph by 
nodes and edges. The multivariate Gaussian 
JPD function can take different forms in which dependencies between the variables 
are described by different types of parameters. Hence, one might build 
various PGGMs that could encode the multivariate Gaussian JPD function.
We describe in some detail two types of PGGMs, in which 
parameters reflect respectively marginal 
dependencies and general conditional dependencies (marginal 
dependencies are special forms of conditional dependencies).

\subsection*{Probabilistic CN models}
The best-known representation of the Gaussian JPD function is in terms of 
marginal dependencies, {\it i.e.}, dependencies of the form $X_i,X_j|\emptyset$
as present in the covariance matrix $\bm{\Sigma}$. Let $\X$ be a $N$-dimensional 
multivariate Gaussian variable then its probability density function 
$P(\X)$ is given by:
\begin{equation}
	\label{eq:mvg-cov}
	P(\X) = (2\pi)^{-N/2}\det(\bm{\Sigma})^{-1/2}\exp \{-1/2(\X-\bm{\mu})^\top\bm{\Sigma}^{-1}(\X-\bm{\mu})\},
\end{equation}
where $\bm{\mu}$ is the $N$-dimensional mean vector and $\bm{\Sigma}$ the 
$N \times N$ covariance matrix.

The corresponding PGGM of the JPD function in equation~(\ref{eq:mvg-cov}) is the 
Probabilistic CN model, {\it i.e.}, the probabilistic 
model that arises from a CN graph. The model is built from the 
network constrained sample matrix $\bm{\Sigma}_\tau$, {\it i.e.} the sample covariance 
matrix $\bm{\Sigma}$ for which all $\Sigma_{ij}$ such that $\Sigma_{ij} \leq \tau$ are 
removed. $\bm{\Sigma}_\tau$ cannot be used directly as an estimator for $\bm{\Sigma}$ 
in equation~(\ref{eq:mvg-cov}) because $\bm{\Sigma}_\tau$ is, in general, not positive 
semi-definite. Instead of $\bm{\Sigma}_\tau$ we use the matrix 
$\bm{\Sigma^F}_\tau$, which is
the positive semi-definite matrix closest to $\bm{\Sigma}_\tau$.  The matrix 
$\bm{\Sigma^F}_\tau$ minimizes the distance to $\bm{\Sigma}_\tau$, 
$||\bm{S}_\tau - \bm{\Sigma^F}_\tau||_F$, 
with the Frobenius norm $||\bm{A}||_F = \sum_{j, k} A_{i,j}^2$ and 
can be computed by using the Higham's
algorithm~\cite{higham_computing_1988}, available 
in the R-package corpcor.

\subsection*{Probabilistic BN models}
Alternatively, the $P(\X)$ in equation~(\ref{eq:mvg-cov}) can be 
characterized with conditional dependencies of the form $X_i|{\cal S}$ with 
${\cal S}\subseteq \mathbf{X}$. The representation of the JPD is then a product of 
conditional probability densities:
\begin{equation}
	\label{eq:mvg-fac} 
	\Prob(X_1,\dots,X_N) = \prod_{i=1}^{N} \Prob_i(\XPi)
\end{equation}
with
\begin{equation}
	\label{eq:mvg-loc} 
	\Prob(\XPi) \sim {\cal N}\left(\mu_i + 
	\sum_{j|X_j \in \Pi_{X_i}}\beta_{ij}(X_j-\mu_j), \; \nu_i\right)
\end{equation}
whenever the set of random variables $\{\XPi\}_{i\in N}$ is 
independent\cite{shachter_gaussian_1989}. In this representation ${\cal N}$ 
is the normal distribution,
$\mu_i$ is the unconditional mean of $X_i$, $\nu_i$ is the conditional 
variance of $X_i$ given the set $\Pi_{X_i}$ and $\beta_{ij}$ is the 
regression coefficient of $X_j$, when $X_i$ is regressed on $\Pi_{X_i}$. 
We call $\Pi_{X_i}$ the parentset of variable $X_i$.

The corresponding PGGM in this case is the Probabilistic BN model. 
The graph of a BN model is a $\DAG$ encoding the 
corresponding probability distribution as in equation~(\ref{eq:mvg-fac}). Each node 
corresponds to a variable $X_i \in \X$, the presence of an arc 
$X_j \rightarrow X_i$ implies the presence of the factor 
$\Prob_i(X_i|\dots X_j \dots )$ in $\Prob(\X)$, and thus conditional 
dependence of $X_i$ and $X_j$. Moreover, the absence of an arc between 
$X_i$ and $X_j$ in the graph implies the absence of the factors 
$\Prob_i(X_i|\dots X_j \dots )$ or $\Prob_j(X_j|\dots X_i \dots )$ 
in $\Prob(\X)$ and, thus, the existence of a set of variables   
${\cal S} \subseteq \X\backslash\{X_i,X_j\}$ that makes $X_i$ and $X_j$  
conditionally independent in 
probability~\cite{Koller:2009:PGM:1795555,castillo_expert_1997}.

The structure of the BN identifies the parentset $\Pi_{X_i}$ 
in equation~(\ref{eq:mvg-fac}). With this structure available, one easily learns 
the corresponding parameter set $(\beta,\nu)$; in our case parameters 
$\beta_{ij}$ and $\nu_i$ are a maximum likelihood fit of the linear 
regression of $X_i$ on its parentset $\Pi_{X_i}$. We use the appropriate 
function in the R-package bnlearn  \cite{scutari_learning_2010}.
The challenge of learning the graph structure is explained in Methods 
Section Learning BNs Structure From Data.

\subsection*{Log-likelihood definition and calculation}
The likelihood of the data $\D$, given a model $\M$ is the density of the 
data under the given model $\M$: $\Prob(\D\given\M)$. For discrete density 
functions the likelihood of the data equals the probability of the data 
under the model. The likelihood is almost always simplified by taking the natural 
logarithm; continuous likelihood values are typically small and 
differentiation of the likelihood function (with the purpose of a 
maximum likelihood search) is often hard. Log-likelihood values can be 
interpreted equally when the expression is used for model comparison 
and maximum likelihood search as the natural logarithm is a 
monotonically increasing function. 

In the following we explain the calculation of the log-likelihood 
$\LL(\D|\M) = \log P(\D|\M)$ for a PGM ($\M = PGM$) for a dataset $\D$ 
formed by n independent data realizations 
$\bm{\D}_k$, $k \in \{1, \dots, n\}$, of the 
$N$-dimensional random vector $\X$, with $\bm{\D}_k = \{d^k_1\dots d^k_N\}$ 
and $d^k_i$ the $k$-th realization of variable $X_i \in \X$. We have
\begin{align}
\log \Prob(\D\given PGM)& = \log \Prob(\bm{\D_1}, \dots, \bm{\D_n}\given PGM) = 
\log \prod_{k=1}^{n}\Prob(\bm{\D}_k\given PGM) \nonumber \\ 
&=  \sum_{k=1}^{n}\log\Prob(\bm{\D}_k\given PGM) = \sum_{k=1}^{n}\log\Prob_{PGM}(\bm{\D}_k) 
\end{align} 
with $P_{PGM}$ the probability density function as modelled 
by the corresponding PGM
with a Gaussian multivariate probability. 
In this work we considered two types of PGMs, correlation and Bayesian 
PGMs, deduced from CNs and BNs graphs, respectively. In the case of correlation 
PGMs, from equation~(\ref{eq:mvg-cov}), we get:
\begin{align}
\LL_\mathrm{CN}(\D\given PGM_\mathrm{CN}) & =  
\sum_{k=1}^{n}\log\Prob(\bm{\D}_k\given PGM_\mathrm{CN}) \nonumber \\
& = \sum_{k=1}^{n}\log\{(2\pi)^{-N/2} \det(\bm{\Sigma^F_\tau})^{-1/2} 
\exp[-1/2(\bm{\D}_k-\bm{\mu})^\top(\bm{\Sigma^F}_\tau)^{-1}(\bm{\D}_k-\bm{\mu})]\}.
\label{eq:loglik_cor}
\end{align}
Entries in the sum (7) are evaluations of the multivariate normal 
density function and executed with the R-package mvtnorm. 
\cite{computation_genz_2009}.

In the case of a PGGM given by a BN, from equation~(\ref{eq:mvg-fac}), we have
\begin{align}
\LL_{BN}(\D\given PGM_{BN}) &= \sum_{k=1}^{n}\log\Prob(\bm{\D_k}\given PGM_{BN}) \nonumber \\& =
\sum_{k=1}^{n}\log\prod_{i=1}^{N} \Prob_i(X_i = d^k_i \given \Pi_{X_i} = 
d^k_{\Pi_{X_i}}) \nonumber \\
&= \sum_{k=1}^{n}\sum_{i=1}^{N} \log\Prob_i(X_i = d^k_i \given \Pi_{X_i} = 
d^k_{\Pi_{X_i}}),  
\label{eq:loglik_bay}
\end{align}
where $d^k_{\Pi_{X_i}}$ is a subset of $\bm{\D}_k$ containing the $k$-th data 
realization of the parentset $\Pi_{X_i}$ of $X_i$. From equation~(\ref{eq:mvg-loc}) 
we know that the conditional univariate densities in the sum in equation~(\ref{eq:loglik_bay}) 
are univariate normal and we execute them with the basic R-package stats.

\bibliography{bibscirep}



\section*{Acknowledgements}

Financial support from Agencia Estatal de 
Investigaci\'on (Spain) and FEDER (EU) is kindly acknowledged; 
JMG and CEG from Project 
MULTI-SDM (CGL2015-66583-R); JML, DP, and MAR 
from Project VACOSCAD (FIS2016-74957-P).


\section*{Additional information}


\textbf{Competing interests} The authors declare that they have no
competing interests.
\textbf{Correspondence} Correspondence and requests for materials
should be addressed to CEG.~(email: catharina.graafland@unican.es).





\end{document}


\maketitle
\thispagestyle{empty}
%
%

\begin{figure}
	\begin{center}
		\includegraphics[width=\textwidth]{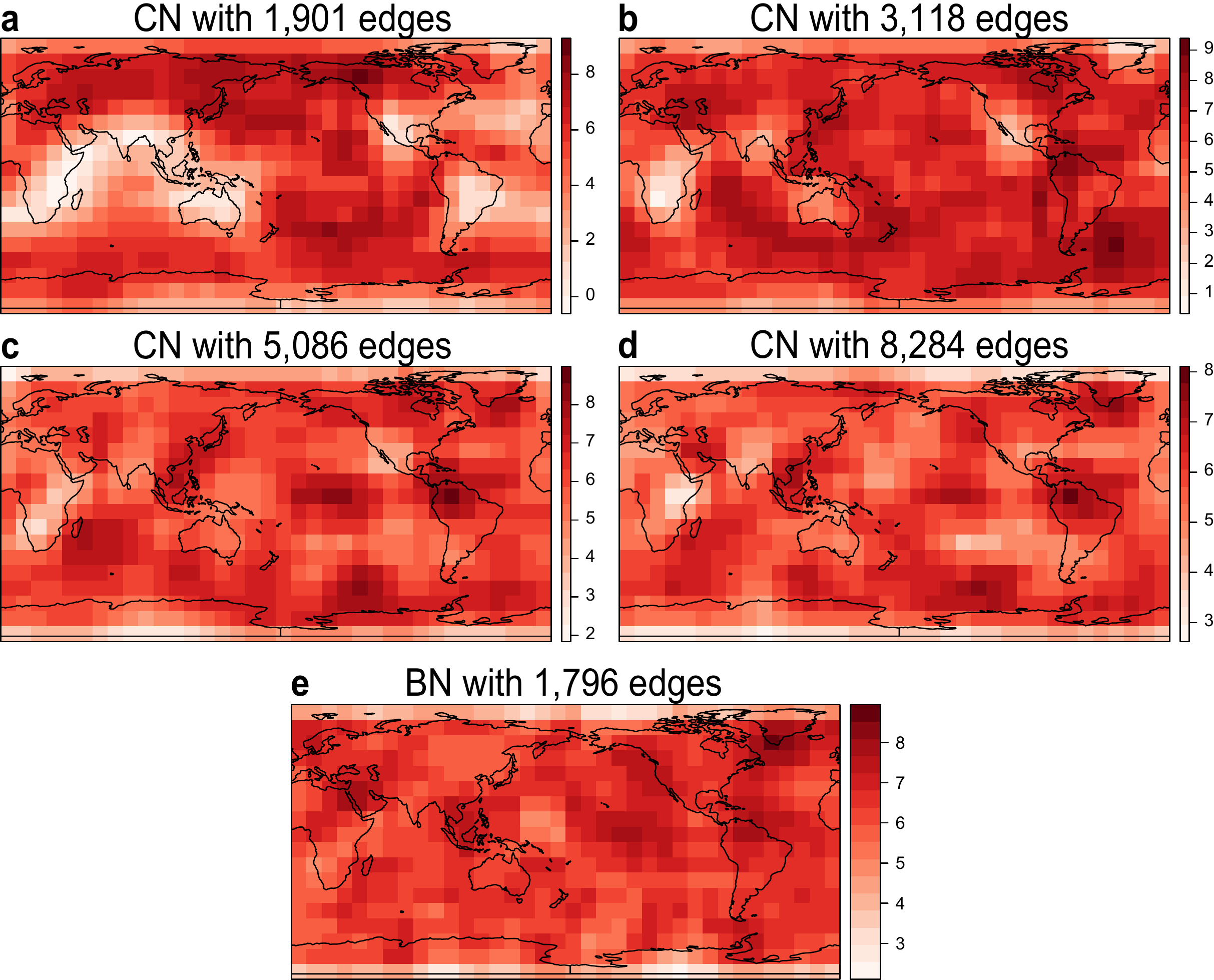}
		\caption{Maps for Correlation Networks (CNs) of size (\textbf{a}) 1,901, (\textbf{b}) 3,118, (\textbf{c}) 5,086 and (\textbf{d}) 8,284 and (\textbf{e}) Bayesian Network of size 1,796 in which gridboxes are coloured in function of their betweenness values. Raw betweenness values $BC_i$ are transformed to $\ln(1+BC_i)$ and every gridbox presents the mean of the betweenness values of its direct neighbors for visualization purposes. CN maps (\textbf{a})-(\textbf{d}) show alternation in the assignation of regions with high and low betweenness. The alternation between the CN maps of size (\textbf{c}) 5,086 and (\textbf{d}) 8,284 is small and one deducts a more ore less established pattern. The white boxes (zero betweenness) in CNs of size (\textbf{a}) 1,901 and (\textbf{b}) 3,118 mostly indicate variables that are unconnected to the network (see largest connected component size in Supplementary Figures 2 and 3). The white boxes in other maps indicate variables that do not belong to any geodesic. CN maps (\textbf{c}) and (\textbf{d}) share high betweenness regions in the mid-east Pacific Ocean, the Northern part of South America together with the Caribbeans, the South-West part of the Indian Ocean, the Philippines and the Chinese Sea and part of the North Atlantic Ocean against Greenland. They also share low betweenness regions for the mid-west Pacific Ocean, the west Pacific Ocean against Mexico, the Eastern part of the Indian Ocean against Australia and South-East part of the Pacific Ocean on the height of Chili. 
			The BN map in (\textbf{e}) coincides on the above mentioned high betweenness regions. The pattern of the BN is little more flattened due to a lower value of the clustering coefficient as displayed in Figure 3.}
		\label{fig:betweeness}
	\end{center}
\end{figure}

\begin{figure}
	\begin{center}
		\includegraphics[width=\textwidth]{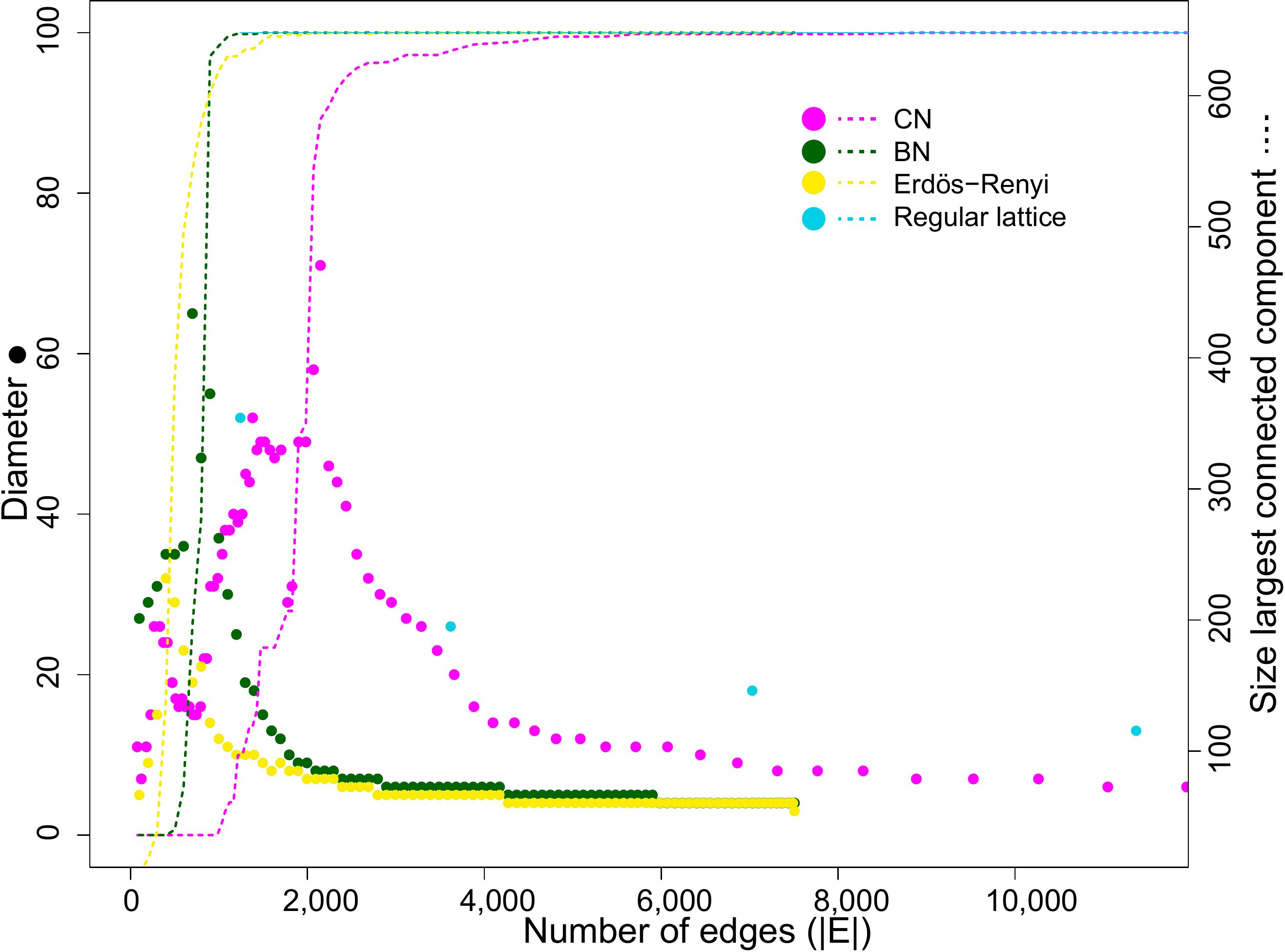}
		\caption{The diameter of a network is the length of the longest geodesic (shortest path) in the network. For a graph that is not fully connected we define the diameter as the length of the longest geodesic of the Largest Connected Component (LCC). The figure displays the diameter (left y-axis, dots) and the size of the largest connected component (right y-axis, dashed lines) versus the number of edges in Correlation Networks (CNs; magenta), Bayesian Networks (BNs; green),  Erd\"os-Renyi random graphs (ERs; yellow) and Regular lattices (RLs; blue).  ERs are random graphs in which every arc is included with probability $2|E|/(N(N-1))$ with $N$ the total number of variables ($N = 648$) and $|E|$ the number of edges in the graph. RLs are deterministic graphs and augment locally. The smallest RL corresponds to a network in which all variables are connected with their direct neighbours in a $36 \times 18$ rectangular grid (order 1 connection). The second RL corresponds to a network in which nodes are connected with direct neighbours and neighbours of direct neighbours (order 2 connection). And so on.\\
			A network is fully connected if the size of the LCC equals $N = 648$. All RLs are fully connected by construction. BNs, ERs and CNs are fully connected at sizes $|E| =$ 1,000, 1,900 and 5,750, respectively.\\
			The maximum diameter value for ERs is 33 at size 500, is 65 for BNs at size 500 and is 70 for CNs at size 2,200. All maximum diameter values are found when the network is almost fully connected. BNs and CNs have similar maxima, but as BNs connect earlier, the size of the BN that yields the maximum diameter is four times smaller than that of the CN. BN diameter values first tend towards those of a local lattice (at size 500), but then tend to efficient ER values (around size 2,000). CN diameter values increase much slower during the network construction process. Values are similar to local lattice structure up to 3,500 edges and do not approach the efficient ER structure up to networks of 8,000 edges and above.}
			\label{fig:diameter}
	\end{center}
\end{figure}

\begin{figure}
	\begin{center}
		\includegraphics[width=\textwidth]{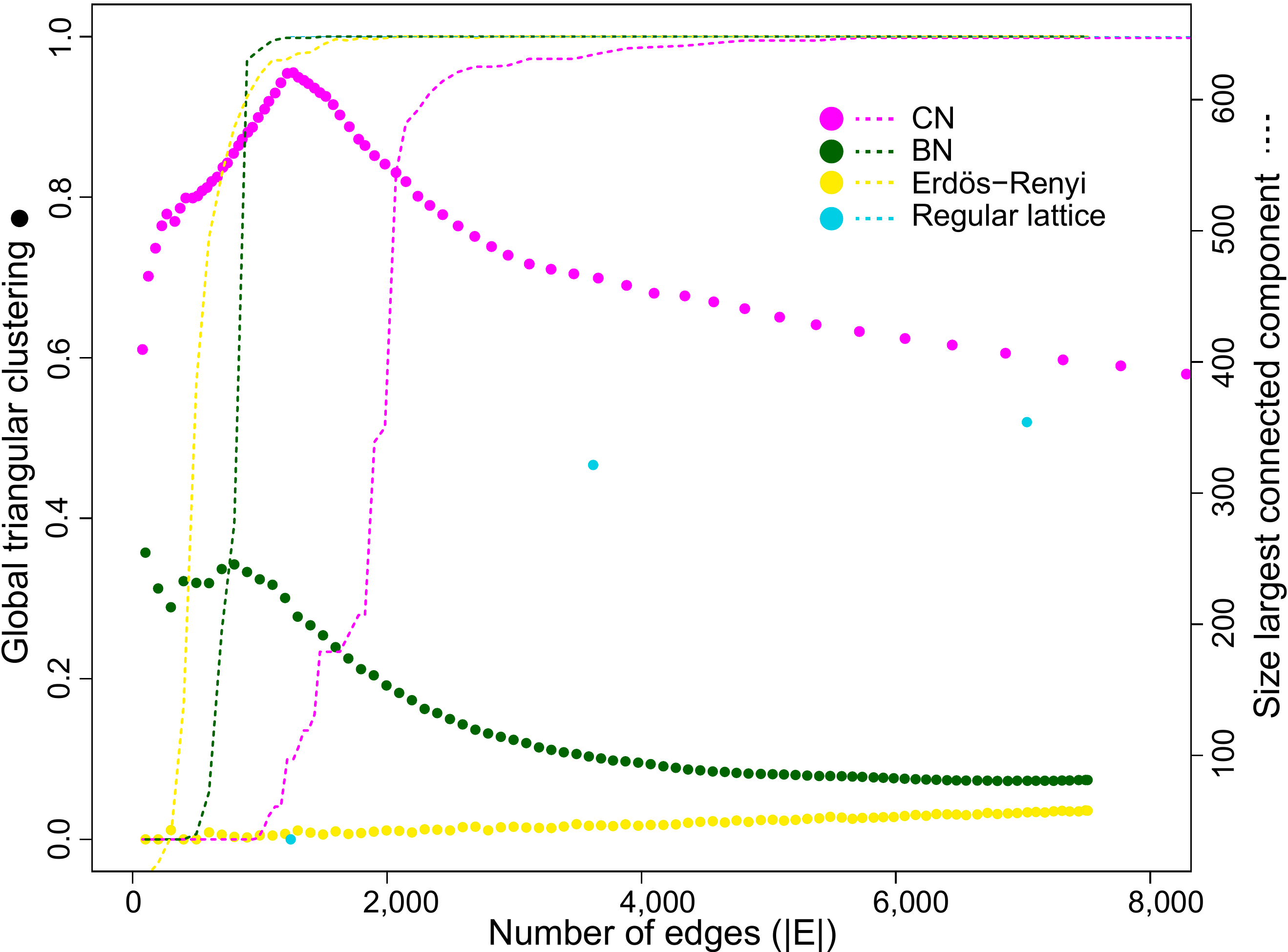}
		\caption{The global triangular clustering of a network is the ratio between the number of triangles and the total number of possible triangles that could exist: 
			$\text{clustering} = \text{number of closed paths of length two} \div \text{number of paths of length two}.$ The figure displays the clustering (left y-axis, dots) and size of the Largest Connected Component (LCC, right y-axis, dashed lines) versus number of edges in Correlation Networks (CNs; magenta), Bayesian Networks (BNs; green),  Erd\"os-Renyi random graphs (ERs; yellow) and Regular lattices (RLs; blue). ERs are random graphs in which every arc is included with probability $2|E|/(N(N-1))$ with $N$ the total number of variables ($N = 648$) and $|E|$ the number of edges in the graph. RLs are deterministic graphs and augment locally. The smallest RL corresponds to a network in which all variables are connected with their direct neighbors (rectangular grid; order 1 connection). The second RL corresponds to a network in which nodes are connected with direct neighbors and neighbors of direct neighbors (order 2 connection). And so on.\\
			ERs do not possess intrinsic clustering structure, and the observed minimal grow of clustering values is only due to network saturation. 
			RLs do have a local clustering structure. This contribution to the clustering coefficients is totally local by definition. Clustering coefficients grow in function of size.
			CNs of all sizes posses high clustering values. A peak value of almost 1 is obtained around 1,700 edges, even in the correlation range with relatively more large distance links clustering values remain high; clustering in CNs occurs at local and global scale (see 
			Figure 1(\textbf{b}-\textbf{c})). 
			BNs posses relative low clustering coefficients when compared with CNs of similar size, however the values are significantly higher then the coefficients of ERs that do not posses clustering structure at all. The peak value 0.35 ($\approx$ one out three connected triples is triangular) belongs to a graph that contains around 500 edges. This graph is almost fully connected, but the structure is still local.
			BNs with more edges have lower clustering coefficients. The BN of 1,796 edges (see corresponding network in Figure 1(\textbf{c}) has a clustering coefficient of 0.2; this BN does exhibit clustering, but the long range edges do not contribute positively to the value of the clustering coefficient as the connection between two locally clustered regions is captured with a minimal amount of long distant edges, instead of an redundant edge bundel as is the case for CNs.}
			\label{fig:clustering}
	\end{center}
\end{figure}